\documentclass[conference,a4paper]{IEEEtran}
\usepackage[english]{babel}
\usepackage{amsmath,amssymb,amsfonts,mathrsfs,amsthm,dsfont,bm,bbm,bbding}

\usepackage[hidelinks,breaklinks=true]{hyperref}
\DeclareMathOperator{\Tr}{Tr}

\usepackage{mathtools}
\mathtoolsset{showonlyrefs=true}
\usepackage{enumitem}
\usepackage{tikz}
\usetikzlibrary{calc,positioning}
\usepackage{adjustbox}

\usepackage[caption=false]{subfig}

\begin{document}

\title{Non-Causal Computation}

\author{
  \IEEEauthorblockN{\"Amin Baumeler and Stefan Wolf}
  \IEEEauthorblockA{Faculty of Informatics, 
    Universit\`a della Svizzera italiana, 
    6900 Lugano, Switzerland\\
    Facolt\`{a} indipendente di Gandria, 
    6978 Gandria, Switzerland\\
    Email: \{baumea,wolfs\}@usi.ch} 
}

\maketitle

\begin{abstract}
	Computation models such as circuits describe sequences of computation steps that are carried out {\em one after the other}. 
	In other words, algorithm design is traditionally subject to the restriction imposed by a fixed causal order. 
	We address a novel computing paradigm beyond quantum computing, replacing this assumption by mere logical consistency: We study {\em non-causal circuits}, where a 
	fixed time structure {\em within\/} a gate is locally assumed whilst the global causal structure {\em between\/} the gates is dropped. 
	We present examples of logically consistent non-causal circuits outperforming all causal ones; they imply that suppressing loops 
	entirely is more restrictive than just avoiding the contradictions they can give rise to. That fact is already known for correlations 
	as well as for communication, and we here extend it to {\em computation}. 
\end{abstract}

\section{Introduction}
Computations, understood as realized through Turing machines, billiard or ballistic computers~\cite{Fredkin:1982bt}, circuits, lists of computer instructions, or otherwise, are often designed to have a linear, {\em i.e.}, causal,  time flow:
After a fundamental operation is carried out, the program counter moves to the next operation, and so forth.
Surely, this is in agreement with our everyday experience; after you finish to read this sentence, you continue to the next (hopefully), or do something else (in that case: goodbye!).
{\em What sorts of computation become admissible if one drops the assumption of a linear time flow and reduces it to mere logical consistency?}
One could imagine that a linear time flow restricts computation strictly {\em beyond what would be allowed for the purely logical point of view}.
Indeed, we show this to be true.
If the assumption of a linear time flow is dropped, a variable of the computational device could depend on ``past'' as well as ``future'' computation steps.
Such a dependence can be interpreted as {\em loops\/} in the time flow, {\it e.g.,} generated by a closed timelike curve~\cite{Echeverria:1991ko}.
There are two fundamental issues that might make loops logically inconsistent.
One is the liability to the {\em grandfather antinomy}.
In a loop-like information flow, multiple contradicting values could potentially be assigned to a variable --- the variable is {\em overdetermined}.
The other issue is {\em underdetermination}:
A~variable could take multiple consistent values, yet, the model of computation {\em cannot predict\/} which actual value it takes.
This underdetermination is also known as the {\em information antinomy}.
To overcome both issues, we restrict ourselves to models of computation where the assumption of a linear time flow is dropped and replaced by the assumption of {\em logical consistency}:
All variables are neither overdetermined nor underdetermined.
We call such models of computation {\em non-causal}.
Our main result is that non-causal models of computation are {\em strictly\/} more powerful than the traditional, causal ones.
Therefore, causality is a stronger assumption than logical consistency in the context of computation.
Similar results are also known with respect to {\em quantum\/} computation~\cite{Chiribella:2012jg,Colnaghi:2012dv,Chiribella:2013bk,Araujo:2014kf,Procopio:2015iw}, correlations~\cite{Oreshkov:2012uh,Chiribella:2013bk,Baumeler:2014cw,Baumeler:2016jt,Branciard:2016bt} as well as communication~\cite{Feix:2015fb}.
As we will shown later, such circuits are ``programmed'' by introducing a {\em contradiction\/} if an {\em undesired\/} result is found.
This is like guessing the solution to a problem and killing the own grandfather in the event that the guess was wrong (similar to ``quantum suicide''~\cite{Tegmark1998} or ``anthropic computing''~\cite{Aaronson2005gusetcolumn}).

The article is structured as follows.
First, we discuss the assumption of logical consistency in more depth, then we describe a non-causal circuit model of computation and give a few examples of problems that can be solved more efficiently.
We continue by describing other non-causal models of computations: the non-causal Turing machine and non-causal billiard computer.
We conclude by showing how to efficiently find a satisfying assignment to a SAT formula if the number of satisfying assignments is {\em previously known}.

\section{Logical consistency}
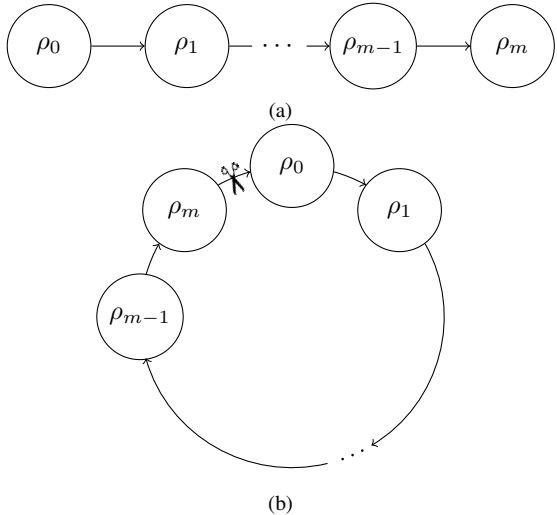
\begin{figure}
	\centering
	\subfloat[\label{fig:causaldep}]{
		\begin{tikzpicture}
			\node[draw,circle,minimum size=1.1cm] (t0) {$\rho_0$};
			\node[draw,circle,minimum size=1.1cm,right=0.7cm of t0] (t1) {$\rho_1$};
			\draw[->] (t0) -- (t1);
			\node[right=0.3cm of t1] (t2) {$\cdots$};
			\draw[-] (t1) -- (t2);
			\node[draw,circle,minimum size=1.1cm,right=0.3cm of t2] (t3) {$\rho_{m-1}$};
			\draw[->] (t2) -- (t3);
			\node[draw,circle,minimum size=1.1cm,right=0.7cm of t3] (t4) {$\rho_m$};
			\draw[->] (t3) -- (t4);
		\end{tikzpicture}
	}
	\quad
	\subfloat[\label{fig:noncausaldep}]{
		\begin{tikzpicture}
			\node (C) {};
			\node[draw,circle,minimum size=1.1cm,shift=(90:2cm)] (t0) at (C.center) {$\rho_0$};
			\node[draw,circle,minimum size=1.1cm,shift=(45:2cm)] (t1) at (C.center) {$\rho_1$};
			\node[draw,circle,minimum size=1.1cm,shift=(3*45:2cm)] (tm) at (C.center) {$\rho_m$};
			\node[draw,circle,minimum size=1.1cm,shift=(4*45:2cm)] (tmm) at (C.center) {$\rho_{m-1}$};
			\draw[->] (C.center)++(90-16:2cm) arc (90-16:45+16:2cm);
			\draw[->] (C.center)++(4*45-16:2cm) arc (4*45-16:3*45+16:2cm);
			\draw[->] (C.center)++(3*45-16:2cm) arc (3*45-16:2*45+16:2cm);
			\draw[line cap=round,line width=1pt,dash pattern=on 0pt off 4pt] (C.center)++(292.5+5:2cm) arc (292.5+5:292.5-5:2cm);
			\draw[->] (C.center)++(360+45-16:2cm) arc (360+45-16:13*360/16+9:2cm);
			\draw[->] (C.center)++(13*360/16-9:2cm) arc (13*360/16-9:4*45+16:2cm);
			\node[rotate=112.5,shift=(0:2cm)] (scissors) at (C.center) {\ScissorLeft};
		\end{tikzpicture}
	}
	\caption{Causal and non-causal computation. The arrows point in direction of computation. (a) The values that are assigned to the variables of a computational model at time~$t$ depend on~$\rho_{t-1}$. (b) Cyclic dependencies of the values that are assigned to the variables at different steps during the computation.}
	\label{fig:dep}
\end{figure}

Let~$\rho_t$ be the ensemble of all variables (also called {\em state}) of a~computational model at a time~$t$.
In general,~$\rho_t$ depends on~$\rho_{t-1},\rho_{t-2},\dots$.
Without loss of generality, assume that~$\rho_{t}$ depends on~$\rho_{t-1}$ only, {\em i.e.}, the computation is described by a~Markov chain.
These dependencies are depicted in Figure~\ref{fig:causaldep}.
In a non-causal model, however, the values that are assigned to the variables at time~$t$ could in principle depend on ``future'' time-steps, {\em e.g.}, the assignment~$\rho_0$ could depend on~$\rho_m$, which results in a Markovian ``bracelet'' or circle (see Figure~\ref{fig:noncausaldep}).

A computational model is {\em not overdetermined\/} if and only if the values that are assigned to the variables do not contradict each other.
This is equivalent to the existence of a fixed point~\cite{Baumeler:2015te} of the Markov chain that results from cutting the ``bracelet'' at an arbitrary position (see Figure~\ref{fig:noncausaldep}).
Let~$f$ be a function that describes the behaviour of this Markov chain.
Then, the computational model is {\em not overdetermined\/} if and only if~$\exists x:f(x)=x$.

A computational model is {\em not underdetermined\/} if and only if there exists at most one fixed point~\cite{Baumeler:2015te}:
\begin{align}
	|\{x\,|\,x=f(x)\}| \leq 1
	\,.
\end{align}

Logical consistency is identified~\cite{Baumeler:2015te} with no overdetermination and no underdetermination, {\it i.e.}, the existence of a {\em unique\/} fixed point:
\begin{align}
	\exists !x:f(x)=x
	\,.
\end{align}

\section{Non-causal circuit model}
A circuit consists of gates that are interconnected with wires.
In the traditional circuit model, back-connections, {\em i.e.}, a cyclic path through a graph where gates are identified with nodes and wires are identified with edges, are either forbidden or interpreted as {\em feedback\/} channels.
An example of a feedback channel is an autopilot system in an aircraft that, depending on the measured altitude, adjusts the rudder and the power setting to maintain the desired altitude, at the same time avoiding a stall.
Here, we interpret back-connections or loops differently.
Whilst in the above scenario the feedback gets introduced at a {\em later\/} point in the computation, the back-action in a non-causal circuit effects the system at an {\em earlier\/} point.
Such a back-action can be interpreted as acting into the past.
Another interpretation is that every gate has its own time (clock), but no global time is assumed --- this interpretation stems from the studies of correlations without causal order~\cite{Oreshkov:2012uh,Chiribella:2013bk}.
Such an interpretation might be more pleasing: Here, ``earlier'' is understood {\em logically}, and the assumption of a global causal order is simply replaced by logical consistency.

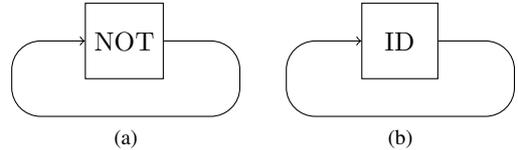
\begin{figure}
	\centering
	\subfloat[\label{fig:linccirc}]{
		\begin{tikzpicture}
			\node[draw,minimum width=1cm,minimum height=1cm] (G) {$\mathrm{NOT}$};
			\draw[rounded corners=0.382cm,->] (G.east) -- ++(1,0) -- ++(0,-1) -- ++(-3,0) -- ++(0,1) -- (G.west);
		\end{tikzpicture}
	}
	\quad
	\subfloat[\label{fig:iinccirc}]{
		\begin{tikzpicture}
			\node[draw,minimum width=1cm,minimum height=1cm] (G) {$\mathrm{ID}$};
			\draw[rounded corners=0.382cm,->] (G.east) -- ++(1,0) -- ++(0,-1) -- ++(-3,0) -- ++(0,1) -- (G.west);
		\end{tikzpicture}
	}
	\caption{(a) Overdetermined circuit: The bit~$0$ is mapped to~$1$ and {\it vice versa\/}, {\it i.e.}, there is no consistent assignment of a value that travels on the wire. (b) Information antinomy: Both~$0$ and~$1$ could potentially travel on the wire, yet the circuit does {\em not specify\/} which.}
	\label{fig:inccirc}
\end{figure}

A {\em non-causal\/} circuit consists of gates that can be interconnected arbitrarily by wires, as long as the circuit as a whole remains logically consistent.
An example of a circuit that is overdetermined and an example of a circuit that leads to the information antinomy (underedetermined) are given in Figure~\ref{fig:inccirc}.

We model a gate~$G$ by a Markov matrix~$\hat G$ with~$0$-$1$ entries.
Without loss of generality, assume that the input and output dimension of a gate are equal.
The Markov matrix of the~$\mathrm{ID}$ gate on a single bit~(see Figure~\ref{fig:iinccirc}) is
\begin{align}
	\mathds{1}=\begin{pmatrix}
		1&0\\
		0&1
	\end{pmatrix}
	\,,
\end{align}
and the Markov matrix of the~$\mathrm{NOT}$ gate on a single bit~(see Figure~\ref{fig:linccirc}) is
\begin{align}
	\hat N=\begin{pmatrix}
		0&1\\
		1&0
	\end{pmatrix}
	\,.
\end{align}
Values are modeled by vectors, {\em e.g.}, in a binary setting, the value~$0$ is represented by the vector~$(1,0)^T$ and the value~$1$ is represented by the vector~$(0,1)^T$.
In general, an~$n$-dimensional variable with value~$i$ is modeled by the~$n$-dimensional vector~$\bm{i}$ with a~1 at position~$i$, and where all other entries are~$0$.
A~gate is applied to a value via the matrix-vector multiplication, {\em i.e.},~the output of~$G$ on input~$a$ is~$\bm{x}=\hat G\bm{a}$.
Let~$F$ and~$G$ be two gates.
The Markov matrix of the parallel composition of both gates is~$\hat F\otimes \hat G$.
They are composed sequentially with a wire that takes the~$d$-dimensional output of~$F$ and forwards it as input to~$G$.
By this, we obtain a new gate~$H=G\circ F$ which represents the sequential composition.
The sequentially composed gate is
\begin{align}
	\hat H=\sum_{v=0}^{d-1} \hat G\bm{v}\bm{v}^T\hat F=\hat G\hat F
	\,.
\end{align}
By using these rules of composition, a {\em causal\/} circuit can always be modeled by a single gate.
A {\em closed\/} circuit is a circuit where all wires are connected to gates on both sides.
Let~$H$ be the gate that describes the composition of all gates for a given causal circuit.
We can transform any such circuit into a closed non-causal circuit by connecting all outputs from~$H$ with all inputs to~$H$.
A {\em logically consistent\/} closed circuit is thus a circuit where a {\em unique\/} assignment of a value~$c$ to the looping wire exists:
\begin{align}
	\bm{c}=\hat H\bm{c} \Longleftrightarrow
	\bm{c}^T\hat H\bm{c}=1
	\label{eq:nc}
	\,.
\end{align}
In other words, the described closed circuit is logically consistent if and only if the diagonal of~$\hat H$ consists of~$0$'s with a single~$1$.
The position of the~$1$-entry represents the fixed point and the value~$c$ on the looping wire.
Note that for a given closed circuit, the gate~$H$ is not unique, but might depend on {\em where\/} the ``cut'' is introduced.
An {\em open\/} circuit is a circuit where some wires are not connected to a gate on one side.
Thus, such a circuit has either an input~$a$, an output~$x$, or both.
A logically consistent open circuit, therefore, is a circuit where for {\em any choice\/} of input~$a$, a {\em unique\/} assignment of a value~$c$ to the looping wire and to the output~$x$ exists, such that
\begin{align}
	(\bm{x}\otimes\bm{c})^T\hat H(\bm{a}\otimes\bm{c})=1
	\,,
\end{align}
where the second output from~$H$ is looped to the second input to~$H$.

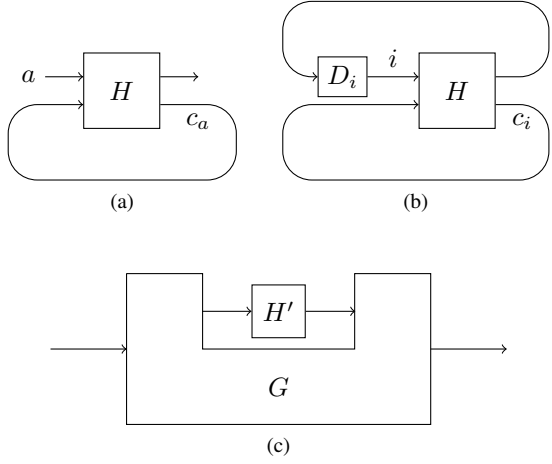
\begin{figure}
	\centering
	\subfloat[\label{fig:open}]{
		\begin{tikzpicture}
			\node[draw,minimum width=1cm,minimum height=1cm] (H) {$H$};
			\draw[->] (H.20) -- ++(0.5,0);
			\draw[<-] (H.160) -- ++(-0.5,0) node[left] {$a$};
			\draw[rounded corners=0.382cm,->] (H.340) -- ++(1,0) node[midway,below] {$c_a$} -- ++(0,-1) -- ++(-3,0) -- ++(0,1) -- (H.200);
		\end{tikzpicture}
	}
	\quad
	\subfloat[\label{fig:close}]{
		\begin{tikzpicture}
			\node[draw,minimum width=1cm,minimum height=1cm] (H) {$H$};
			\node[draw,minimum width=0.5cm,minimum height=0.5cm,xshift=-1cm] (D) at (H.160) {$D_i$};
			\draw[rounded corners=0.382cm,->] (H.340) -- ++(0.7,0) node[midway,below] {$c_i$} -- ++(0,-1) -- ++(-3.5,0) -- ++(0,1) -- (H.200);
			\draw[->] (D.east) -- (H.160) node[midway,above] {$i$};
			\draw[rounded corners=0.382cm,->] (H.20) -- ++(0.7,0) -- ++(0,1) -- ++(-3.5,0) -- ++(0,-1) -- (D.west);
		\end{tikzpicture}
	}
	\par\bigskip
	\subfloat[\label{fig:comb}]{
		\begin{tikzpicture}
			\draw[-] (0,0) -- ++(1,0) -- ++(0,-1) -- ++(2,0) -- ++(0,1) -- ++(1,0)
				 -- ++(0,-2) -- ++(-4,0) -- (0,0);
			 \node[draw,minimum height=0.7cm,minimum width=0.7cm] (H) at (2,-0.5) {$H'$};
			 \draw[->] (H.east) -- ++(0.65,0);
			 \draw[<-] (H.west) -- ++(-0.65,0);
			 \draw[<-] (0,-1) -- ++(-1,0);
			 \draw[->] (4,-1) -- ++(1,0);
			 \node[below=0.4cm of H] {$G$};
		\end{tikzpicture}
	}
	\caption{(a) Open circuit~$\mathcal{C}$ with input~$a$. (b) Closed circuit~$\mathcal{C}_i$ with~$a=i\rightarrow  c_a=c_i$. (c) The big box represents a comb that transforms a gate ($H'$) to a new gate, the composition.}
	\label{fig:openclose}
\end{figure}

Let~$c_a$ be the value on the looping wire of a logically consistent open circuit~$\mathcal{C}$ with input~$a$.
We can transform~$\mathcal{C}$ into a family~$\{\mathcal{C}_i\}_{0\leq i<d}$ of logically consistent {\em closed\/} circuits such that the value on the same looping wire of~$\mathcal{C}_i$ is~$c_i$.
The circuit~$\mathcal{C}_i$ is constructed by attaching the gate
\begin{align}
	\hat D_i=\sum_{v=0}^{d-1}\bm{i}^T\bm{v}
\end{align}
to the input and output wires of~$\mathcal{C}$ (see Figures~\ref{fig:open} and~\ref{fig:close}).
The gate~$D_i$ unconditionally outputs the value~$i$.

There is an ambiguity on which wires are regarded as ``looping.''
We show that two different representations~$H$ and~$H'$ of the same closed non-causal circuit~$\cal{C}$ yield the same computation (the difference between~$H$ and~$H'$ is the identification of the looping wires).
Different~$H$ and~$H'$ that represent the {\em same\/} non-causal circuit~$\cal{C}$ can be written as~$H=Q\circ R$ and~$H'=R\circ Q$.
For~$H$, the looping wires are those that exit~$Q$ and enter~$R$, and for~$H'$, {\it vice versa}.
From Equation~\eqref{eq:nc} we have
\begin{align}
	\exists!c:\bm{c}^T\hat H\bm{c}=\bm{c}^T\hat Q\hat R\bm{c}=\bm{c}^T\hat Q\left( \sum_e\bm{e}\bm{e}^T \right)\hat R\bm{c}=1
	\,.
\end{align}
Since~$R$ is deterministic, the value of~$e$ is uniquely determined.
Thus, we obtain
\begin{align}
	\exists!c:\bm{c}^T\hat Q\bm{e}_*\bm{e}_*^T\hat R\bm{c}=1
	\,,
\end{align}
where~$e_*$ is the specific value on the wire exiting~$R$ and entering~$Q$.
Conversely,
\begin{align}
	\exists!e':\bm{e}'^T\hat H'\bm{e}'&=\bm{e}'^T\hat R\hat Q\bm{e}'\\
	&=\bm{e}'^T\hat R\left( \sum_{c'}\bm{c}'\bm{c}'^T \right)\hat Q\bm{e}'\\
	&=\bm{e}'^T\hat R\bm{c}'_*\bm{c}'^T_*\hat Q\bm{e}'=1
	\,,
\end{align}
holds.
The only way~$H$ and~$H'$ each have a {\em unique\/} fixed point is with the identification~$e_*=e'$.
Therefore, both representations~$H$ and~$H'$ assign the same values to the wires.
By the above translation from {\em open\/} to {\em closed\/} circuits, we see that the same reasoning can be applied to open circuits.

Above, we considered {\em deterministic\/} Markov processes.
It is natural to extend this model to probabilistic processes, {\em i.e.},~stochastic matrices.
The logical-consistency condition in that case, as studied in Ref.~\cite{Baumeler:2015te}, is
\begin{align}
	\Tr\hat H&=1\,,\label{eq:tr}\\
	\forall i,j:\hat H_{i,j}&\geq 0
	\,,
\end{align}
{\it i.e.}, the diagonal of~$\hat H$ consists of non-negative numbers (probabilities) that add up to~$1$.
Equation~\eqref{eq:tr} can be interpreted as ``the average number of fixed points is 1.''
To see this, we decompose~$H$ as a convex combination of {\em deterministic\/} matrices
\begin{align}
	\hat H=\sum_i p_i \hat H_i
	\,,
\end{align}
where for all~$i$,~$\hat H_i$ is deterministic.
Then, Equation~\eqref{eq:tr} states
\begin{align}
	\Tr\hat H=\sum_i p_i\Tr\hat H_i=1
	\,.
\end{align}
For an arbitrary deterministic matrix~$\hat D$, the expression~$\Tr\hat D$ represents the {\em number\/} of fixed points, with which we arrive at the stated interpretation.

An open non-causal circuit can be represented by a comb~\cite{Chiribella:2013bk}~$G$ which is a higher-order transformation ---~$G$ transforms the gate~$H'$ to a new gate (see Figure~\ref{fig:comb}).
The comb~$G$, for instance, could connect the output from~$H'$ with the input of~$H'$, as long as the composition remains logically consistent.

\section{Computational advantage}
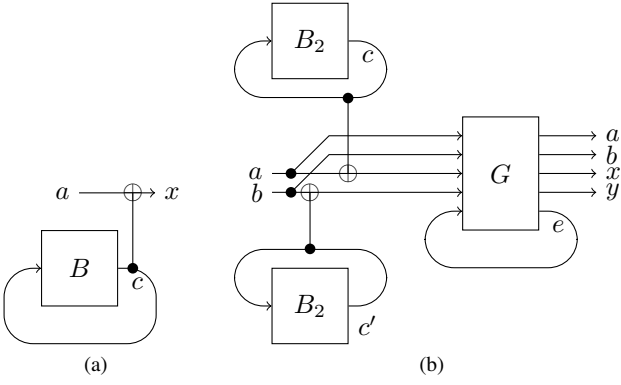
\begin{figure}
	\centering
	\subfloat[\label{fig:bb}]{
		\begin{tikzpicture}
			\node[draw,minimum width=1cm,minimum height=1cm] (B) {$B$};
			\draw[rounded corners=0.382cm,->] (B.east) -- ++(0.5,0) node[midway,below] {$c$} -- ++(0,-1) -- ++(-2,0) -- ++(0,1) -- (B.west);
			\draw[->] (B.west)++(0.5,1) node[left] {$a$} -- ++(1,0) node[right] {$x$};
			\fill (0.7,0) circle [radius=2pt];
			\draw[-] (0.7,0) -- ++(0,1) node {$\oplus$};
		\end{tikzpicture}
	}
	\quad
	\subfloat[\label{fig:2bb}]{
		\begin{tikzpicture}
			\node[draw,minimum width=1cm,minimum height=1cm] (B) {$B_2$};
			\node[draw,minimum width=1cm,minimum height=1cm,below=2.5cm of B] (B2) {$B_2$};
			\draw[rounded corners=0.382cm,->] (B.east) -- ++(0.5,0) node[midway,below] {$c$} -- ++(0,-0.75) -- ++(-2,0) -- ++(0,0.75) -- (B.west);
			\draw[rounded corners=0.382cm,->] (B2.east) -- ++(0.5,0) node[midway,below] {$c'$} -- ++(0,0.75) -- ++(-2,0) -- ++(0,-0.75) -- (B2.west);
			\node[draw,minimum width=1cm,minimum height=1.5cm,below right=1cm and 2cm of B.center] (G) {$G$};
			\coordinate (GE1) at ($ (G.east) + (0,0.5) $);
			\coordinate (GE2) at ($ (G.east) + (0,0.25) $);
			\coordinate (GE3) at ($ (G.east) + (0,0) $);
			\coordinate (GE4) at ($ (G.east) + (0,-0.25) $);
			\coordinate (GE5) at ($ (G.east) + (0,-0.5) $);
			\coordinate (GW1) at ($ (G.west) + (0,0.5) $);
			\coordinate (GW2) at ($ (G.west) + (0,0.25) $);
			\coordinate (GW3) at ($ (G.west) + (0,0) $);
			\coordinate (GW4) at ($ (G.west) + (0,-0.25) $);
			\coordinate (GW5) at ($ (G.west) + (0,-0.5) $);
			\draw[<-] (GW3) -- ++(-1.5,0) node (P1) {$\oplus$} -- ++(-0.75,0) node[fill,circle,minimum size=4pt,inner sep=0pt] (C1) {} -- ++(-0.25,0) node[left] {$a$};
			\draw[->] (GE3) -- ++(0.75,0) node[right] {$x$};
			\draw[<-] (GW4) -- ++(-2,0) node (P2) {$\oplus$} -- ++(-0.25,0) node[fill,circle,minimum size=4pt,inner sep=0pt] (C2) {} -- ++(-0.25,0) node[left] {$b$};
			\draw[->] (GE4) -- ++(0.75,0) node[right] {$y$};
			\draw[rounded corners=0.382cm,->] (GE5) -- ++(0.5,0) node[midway,below] {$e$} -- ++(0,-0.75) -- ++(-2,0) -- ++(0,0.75) -- (GW5);
			\draw[-] (P1.center) -- ++(0,1) node (Q1) {};
			\fill (Q1.center) circle [radius=2pt];
			\draw[-] (P2.center) -- ++(0,-0.75) node (Q2) {};
			\fill (Q2.center) circle [radius=2pt];
			\draw[->] (C1.center) -- ++(0.5,0.5) -- (GW1);
			\draw[->] (C2.center) -- ++(0.5,0.5) -- (GW2);
			\draw[->] (GE1) -- ++(0.75,0) node[right] {$a$};
			\draw[->] (GE2) -- ++(0.75,0) node[right] {$b$};
		\end{tikzpicture}
	}
	\caption{Fixed point search for a black box with one and a black box with two fixed points. (a) The output~$x$ is the fixed point~$c$ added to the input~$a$. (b) Circuit for finding a fixed point for a black box with {\em two\/} fixed points.}
	\label{fig:bbox}
\end{figure}

The logical-consistency requirement forces the value on a looping wire to be the unique fixed point of the transformation.
This can be exploited for {\em finding fixed points\/} of a black box.
Suppose we are given a black box~$B$ that takes (produces) a~$d$-dimensional input (output) and has a {\em unique\/} fixed point~$x$ previously unknown to us.
As a Markov matrix,~$B$ is
\begin{align}
	\hat B=\sum_{i=0}^{d-1} \bm{e}_i\bm{i}^T\,,\quad\text{with }\left|\{i\,|\,e_i=i\}\right|=1
	\,.
\end{align}
Our task is to find the fixed point~$x$ in as few queries as possible.
If we solve this task with a causal circuit, then, in the worst case,~$d-1$ queries are needed.
In contrast, with a non-causal circuit, a {\em single\/} query suffices.
The reason for this is that the black box is queried with the fixed point only.
Any other query would lead to a logical contradiction, and, therefore, does not occur.
For that purpose, we just connect the output of~$B$ with the input of~$B$ and use a second wire to read out the value (see Figure~\ref{fig:bb}).
This circuit is logically consistent because
\begin{align}
	\forall a,\exists !c,x&:(\bm{x}\otimes\bm{c})^T\hat C(\mathds{1}\otimes\hat B)(\bm{a}\otimes\bm{c})\\
	&=(\bm{x}\otimes\bm{c})^T\hat C(\bm{a}\otimes\hat B\bm{c})=1
	\,,
\end{align}
where~$\hat C$ is the CNOT gate and~$\mathds{1}$ is the identity.
This construction, however, works only if~$B$ has a {\em unique\/} fixed point.
Suppose~$B_2$ has {\em two\/} fixed points.
In that case, the circuit from Figure~\ref{fig:2bb} can be used to find both fixed points with two queries.
Additionally to short-cutting the black boxes, we need to introduce a gate~$G$ that ensures a {\em unique\/} fixed point of the whole circuit.
The gate~$G$ works in the following way
\begin{align}
	\hat G=&\sum_{e,c-a<c'-b}(\bm{a}\otimes\bm{b}\otimes\bm{c}\otimes \bm{c'}\otimes \bm{0}) (\bm{a}\otimes\bm{b}\otimes\bm{c}\otimes \bm{c'}\otimes \bm{e})^T+\notag\\
	&\sum_{e,c-a\geq c'-b}(\bm{a}\otimes\bm{b}\otimes\bm{c}\otimes \bm{c'}\otimes \bm{\bar e}) (\bm{a}\otimes\bm{b}\otimes\bm{c}\otimes \bm{c'}\otimes \bm{e})^T
	\,,
\end{align}
where~$e$ is binary,~$\bar e=e\oplus 1$, the addition is carried out modulo~$2$, and~$\bm{0}$ is a~$2$-dimensional vector representing the value~$0$.
In words, if the value~$c$ on the upper wire is less than the value on the lower wire~$c'$, and~$e$ is~$0$, then we get a fixed point on the third wire of~$G$ (variable~$e$ in Figure~\ref{fig:2bb}).
Otherwise, the bit on the third wire gets flipped --- no fixed point.
This guarantees that all loops together have a {\em unique\/} fixed point.
Ironically, the gate~$G$ suppresses certain fixed points on the previous loops by introducing a logical {\em inconsistency\/} at a later point in the circuit. 
This resembles ``anthropic computing''~\cite{Aaronson2005gusetcolumn}, where one guesses the solution to a problem and commits suicide if the guess was wrong --- a recipe to solve NP-complete problems in the relative-state interpretation of quantum mechanics~\cite{Everett1957} and where consciousness follows only those branches where the programmer remains alive.
Such a construction can be used to find the fixed points of a black box with a {\em few\/} fixed points and where the number of fixed points is {\em known}.
For a large number~$n$ of fixed points, {\it e.g.},~$n=d/2$, we can use the probabilistic approach to non-causal circuits.
Let~$B_n$ be a black box with~$n$ fixed points and input and output spaces of dimension~$d$.
The Markov matrix of~$B_n$ is
\begin{align}
	\hat B_n=\sum_{i=0}^{d-1}\bm{e}_i\bm{i}^T\,,\quad\text{with }\left|\{i\,|\,e_i=i\}\right|=n
	\,.
\end{align}
We construct a randomized gate where the average number of fixed points is one:
\begin{align}
	\hat B'=\frac{1}{n}\hat B_n+\frac{n-1}{n}\hat N
	\,,
\end{align}
with
\begin{align}
	\hat N=\sum_{i=0}^{n-1}\bm{\bar i}\bm{i}^T\,,\quad \bar i=i\oplus 1
	\,.
\end{align}
The gate~$\hat N$ can be understood as a~$d$-dimensional generalization of the~$\mathrm{NOT}$ gate for bits:
The input is increased by one modulo~$d$.
Such an~$\hat N$ has {\em no\/} fixed points.
The mixture~$\hat B'$ is logically consistent, because
\begin{align}
	\Tr\left(\frac{1}{n}\hat B_n+\frac{n-1}{n}\hat N\right)=\frac{1}{n}\Tr \hat B_n+\frac{n-1}{n}\Tr \hat N =1
	\,.
\end{align}
This means that we can use the circuit from Figure~\ref{fig:bb} to find a random fixed point of~$B_n$.

We apply these tools to find solutions to instances of search problems with a {\em known\/} number of solutions, and where a guess for a solution can be verified efficiently by a verifier~$V$.
In other words, we can find solutions to NP search problems, yet where the number of solutions to an instance must be known to us in advance.
Note that the following construction does not solve a decision problem, but rather {\em finds\/} the solution.
Suppose an instance~$I$ to a problem~$\Pi$ has a {\em unique\/} solution.
We replace the gate~$B$ of Figure~\ref{fig:bb} with a new gate~$V'$ that acts in the following way: It takes a guess~$c$ for a solution to~$\Pi(I)$ as input, runs~$V$ to verify~$c$.
If~$V$ accepts~$c$, then~$V'$ outputs~$c$, and otherwise,~$V'$ outputs~$c\oplus 1$, where the addition is carried out modulo~$d$.
Such a circuit has a unique fixed point~$c$ which equals the solution of~$\Pi(I)$.
This, for instance, could be applied to a~$\mathrm{SAT}$ formula, where a {\em unique\/} assignment of values to variables exist which make the formula true.

\section{Other non-causal computational models}
We briefly discuss non-causal Turing machines and non-causal billiard computers.
A~Turing machine~$T$ has a~tape, a~read/write head, and an internal state machine.
After every read instruction, the state machine moves to the next internal state, and thereby decides what to write and where to move the head to.
A non-causal Turing machine is a machine where parts of the tape are not ``within time:''
``Future'' (from the head's point of view) {\em write\/} instructions influence ``past'' {\em read\/} instructions.
A~symbol that is written at time~$t$ to position~$j$ could be read at time~$t'<t$ form position~$j$, {\em i.e.}, symbols can be read ``before'' they are written.
This, as other self-referential systems, leads to problems that can be solved if we enforce the condition of logical consistency, as discussed above.
Another issue is that multiple {\em write\/} instructions could {\em overwrite\/} the value on position~$j$.
This leaves open the question what value is read.
We can overcome this issue by running the Turing machine in a reversible fashion and by generating a history tape~\cite{Bennett:1973ko}, where {\em no\/} memory position gets overwritten.
An example of a non-causal Turing machine is where the {\em history\/} tape is non-causal in the sense that symbols can be read ``before'' they are written.

The billiard computer is a model of computation on a~billiard table~\cite{Fredkin:1982bt}.
Before the computation starts, obstacles are placed on the table in such a way that the induced reflections of the balls and the collisions among the balls result in the desired computation.
A non-causal version of a billiard computer is a billiard table where the holes are connected with closed timelike curves (CTCs)~\cite{Echeverria:1991ko} that are logically consistent.
Now, a billiard ball could also collide with its younger self; this introduces a non-causal effect.
Echeverria, Klinkhammer, and Thorne~\cite{Echeverria:1991ko} showed that solutions to CTC-dynamics that are not overdetermined exist.
However, all solutions that they found are underdetermined.
The non-causal circuits presented in this work indicate that also logically consistent non-causal billiard computers are admissible.

\section{Conclusion and open questions}
We show that models of computation, where parts of the output of a computation are (re)used as input to the {\em same\/} computation, are logically possible.
Furthermore, such a model of computation helps to solve certain tasks more efficiently.
The question is how much more powerful this new model of computation is, and whether uncomputable tasks become computable when compared to the standard circuit model.
A strong restriction of the model is that, before one can find a fixed point, one needs to know the number of fixed points.
For instance, if we want to find a satisfying assignment for a SAT formula~$F$ with variables~$x_0,x_1,\dots$, we first need to know the number of satisfying assignments --- otherwise we do not know how to construct the circuit.
Ironically, this means that to solve a SAT problem without any promise, we first need to solve a problem that is believed to be much harder: a \#SAT problem.
One might want to apply the Valiant-Vazirani~\cite{Valiant:1986ks} method to~$F'=F\vee(x_0\wedge x_1\wedge\dots)$ to reduce the number of satisfying assignments to~1.\footnote{The reason why we modify~$F$ to~$F'$ is to guarantee satisfiability.} 
The problem that we are left with is: We do {\em not\/} know whether the output~$F''$ of the Valiant-Vazirani method has a unique satisfying assignment or not --- the reduction is probabilistic.
Therefore, we cannot plug~$F''$ into a circuit, like the one shown in Figure~\ref{fig:bb}, to find the fixed point.

A model of computation similar to but more general than ours is based on Deutsch's~\cite{Deutsch:1991jo} CTCs.
Aaronson and Watrous~\cite{Aaronson:2009dy} showed that the classical special case of Deutsch's model can solve problems in PSPACE efficiently.
However, in Deutsch's model, in contrast to ours, the information antinomy arises.
Deutsch mitigates this issue by defining that the value on the looping wire is the uniform mixture of all solutions.
This introduces a non-linearity into Deutsch's model: The output of a circuit depends non-linearly on the input.
A consequence of this is that --- in the quantum version --- quantum states can be cloned~\cite{Brun:2013jx}.
The model studied here, as it is linear, is not exposed to such consequences.

\section*{Acknowledgments}
We thank Mateus Ara{\'u}jo, Veronika Baumann, Cyril Branciard, {\v C}aslav Brukner, Fabio Costa, Paul Erker, Adrien Feix, Arne Hansen, Alberto Montina, Christopher Portmann, and Benno Salwey for helpful discussions. This work was supported by the Swiss National Science Foundation (SNF), the National Centre of Competence in Research ``Quantum Science and Technology'' (QSIT) and the COST action on Fundamental Problems in Quantum Physics.

\bibliographystyle{IEEEtran}
\bibliography{refs}
\end{document}